\def \inparg{\leftskip = 40pt\rightskip = 40pt}
\def \outparg{\leftskip = 0 pt\rightskip = 0pt}

\def\npb{{Nucl.\ Phys.\ }{\bf B}}
\def\plb{{Phys.\ Lett.\ }{ \bf B}}

\def\prd{{Phys.\ Rev.\ }{\bf D}}

\def\prl{Phys.\ Rev.\ Lett.\ }

\def\Ttil{\tilde T}

\def\frak#1#2{{\textstyle{{#1}\over{#2}}}}

\def\lambdabar{\bar\lambda}
\def\sigmabar{\bar\sigma}

\def\Ttil{\tilde T}

\def\alphadot{\dot\alpha}
\def\betadot{\dot\beta}
\def\deltadot{\dot\delta}

\def\pa{\partial}

\input harvmac
\input epsf
{\nopagenumbers
\line{\hfil LTH 643}
\line{\hfil hep-th/0412009}
\vskip .5in
\centerline{\titlefont One-loop renormalisation of}
\centerline{\titlefont $N=\frak12$ supersymmetric gauge theory}
\vskip 1in
\centerline{\bf I.~Jack, D.R.T.~Jones and L.A.~Worthy}
\bigskip
\centerline{\it Department of Mathematical Sciences,  
University of Liverpool, Liverpool L69 3BX, U.K.}
\vskip .3in
We show that $N=\frak12$ supersymmetric gauge theory is renormalisable
at one loop, 
but only after gauge invariance is restored in a non-trivial fashion. 
\Date{December 2004}}

\newsec{Introduction} There has recently been much interest in theories
defined on non-anti-commutative superspace \ref\seib{N.~Seiberg, JHEP
{\bf 0306} (2003) 010} \ref\araki{T.~Araki, K.~Ito and  A. Ohtsuka,
\plb573 (2003) 209}. Such theories are non-hermitian and turn out to
have only half the supersymmetry of the corresponding ordinary
supersymmetric theory--hence the term ``$N=\frak12$ supersymmetry''. 
These theories are not power-counting  renormalisable but it has been
argued\ref\terash{S. Terashima and J-T Yee, JHEP {\bf 0312} 
(2003) 053}
\nref\gris{M.T.~Grisaru, S.~Penati and  A.~Romagnoni, JHEP {\bf
0308} (2003) 003\semi
R.~Britto and B.~Feng, \prl91 (2003) 201601\semi
A.~Romagnoni, JHEP {\bf 0310} (2003) 016}\nref\lunin{O.~Lunin 
and S.-J. Rey, JHEP  {\bf 0309}
(2003) 045}--\ref\berrey{D.~Berenstein and S.-J.~Rey, \prd68 (2003) 121701}
 that they are in  fact nevertheless
renormalisable, in the sense that only a finite number of additional
terms need to be added to the lagrangian to absorb divergences to all
orders. This is primarily because although the theory contains operators
of dimension five and higher, they are not accompanied by their
hermitian conjugates which would be required to generate divergent
diagrams. This argument does not of course guarantee that the precise 
form of the 
lagrangian will be preserved by renormalisation; nor does the $N=\frak12$ 
supersymmetry, since some terms in the lagrangian are inert under this
symmetry.    
Moreover, the argument also requires (in the gauged case) the
assumption of gauge invariance to rule out some classes of divergent 
structure. We shall show here by explicit calculation that there are
problems with this assumption; even at one loop, at least in the
standard class of gauges, divergent non-gauge-invariant terms are
generated. However, in the case of pure $N=\frak12$
supersymmetry (i.e. no chiral matter) we shall display a divergent field 
redefinition
which miraculously removes the non-gauge-invariant terms and restores 
gauge invariance. The form of the lagrangian
is not quite preserved by renormalisation. However, a slightly modified
version of the original lagrangian (which is still $N=\frak12$
supersymmetric) does have a form preserved under renormalisation. We shall  
try to give
sufficient details of our calculations in order to enable the interested
 reader to check them. The one-loop calculation of the divergences in
pure $N=\frak12$ gauge theory has been performed before \ref\alish{
M.~Alishahiha, A.~Ghodsi and N.~Sadooghi, \npb691
(2004) 111} but we disagree with this result as
we shall explain. The original non-anticommutative theory defined in 
superfields appears to require a $U(N)$ gauge group\araki\terash. However the 
component form of the action is easily adapted to $SU(N)$, as we shall see;
and we shall argue that in addition to being simpler for calculational purposes,
it is only for $SU(N)$ that a form-invariant lagrangian
can be defined.  
We shall therefore assume the gauge group $SU(N)$ so that the basic
commutation relations are (for the fundamental representation): 
\eqn\commrel{ [R^a,R^b]=if^{abc}R^c,\quad
\{R^a,R^b\}=d^{abc}R^c+{1\over N}\delta^{ab},} 
where $d^{abc}$ is totally symmetric.
The action for pure $N=\frak12$ supersymmetric gauge theory is given 
in components by\seib
\eqn\lagran{\eqalign{
 S=&\frak12\int d^4x
\Bigl[-\frak12F^{\mu\nu a}F^a_{\mu\nu}-2i\lambdabar^a\sigmabar^{\mu}
(D_{\mu}\lambda)^a+D^aD^a
-igC^{\mu\nu}d^{abc}F^a_{\mu\nu}\lambdabar^b\lambdabar^c\cr
&+\frak14g^2|C|^2[d^{abe}d^{cde}(\lambdabar^a\lambdabar^b)
(\lambdabar^c\lambdabar^d)
+\frak2Nh(\lambdabar^a\lambdabar^a)(\lambdabar^b\lambdabar^b)]\Bigr]\cr}} 
with gauge coupling $g$, gauge field $A_{\mu}$, gaugino $\lambda$ and with
\eqn\fmunu{\eqalign{
F_{\mu\nu}^a=&\pa_{\mu}A_{\nu}^a-\pa_{\nu}A_{\mu}^a-gf^{abc}A_{\mu}^bA_{\nu}^c
,\cr
D_{\mu}\lambda^a=&\pa_{\mu}\lambda^a-gf^{abc}A_{\mu}^b\lambda^c,\cr}}
and $h=1$. (The $U(N)$ action would have a similar form to Eq.~\lagran,
except that the gauge indices would run over the extra $U(1)$ generator
and we would have $h=0$.)
In deriving Eq.~\lagran\ from the form given in Ref.~\seib\ we are 
assuming that $\lambda=\lambda^aR^a$,
$A_{\mu}=A_{\mu}^aR^a$ so that for instance $\Tr[(\lambdabar\lambdabar)^2]=
\frak18[d^{abe}d^{cde}(\lambdabar^a\lambdabar^b)
(\lambdabar^c\lambdabar^d)
+\frak2N(\lambdabar^a\lambdabar^a)(\lambdabar^b\lambdabar^b)]$; another choice 
of representation for $\lambda$, $A_{\mu}$ would simply result in 
a rescaling of the lagrangian. 
In Eq.~\lagran, $C^{\mu\nu}$ is related to the non-anti-commutativity 
parameter $C^{\alpha\beta}$ by  
\eqn\Cmunu{
C^{\mu\nu}=C^{\alpha\beta}\epsilon_{\beta\gamma}
\sigma^{\mu\nu}_{\alpha}{}^{\gamma},} 
where 
\eqn\sigmunu{\eqalign{
\sigma^{\mu\nu}=&\frak14(\sigma^{\mu}\sigmabar^{\nu}-
\sigma^{\nu}\sigmabar^{\mu}),\cr
\sigmabar^{\mu\nu}=&\frak14(\sigmabar^{\mu}\sigma^{\nu}-
\sigmabar^{\nu}\sigma^{\mu}),\cr }} 
and 
\eqn\Csquar{
|C|^2=C^{\mu\nu}C_{\mu\nu}.} 
Our conventions are in accord with \seib; in particular, 
\eqn\sigid{
\sigma^{\mu}\sigmabar^{\nu}=-\eta^{\mu\nu}+2\sigma^{\mu\nu}.}
Properties of $C$ which follow from
Eq.~\Cmunu\ are  
\eqn\cprop{\eqalign{
C^{\mu\nu}\sigma_{\nu\alpha\betadot}&=C_{\alpha}{}^{\gamma}
\sigma^{\mu}{}_{\gamma\betadot},\cr
C^{\mu\nu}\sigmabar_{\nu}^{\alphadot\beta}&=-C^{\beta}{}_{\gamma}
\sigmabar^{\mu\alphadot\gamma}.\cr}} 

The $N=\frak12$ theory in components is 
manifestly invariant under the standard $SU(N)$ gauge transformations.
Moreover Eq.~\lagran\ is invariant under the $N=\frak12$ supersymmetry 
transformations (adapted to the $SU(N)$ case from Ref.~\seib)
\eqn\susytran{\eqalign{
\delta A^a_{\mu}=&-i\lambdabar^a\sigmabar_{\mu}\epsilon\cr
\delta \lambda^a_{\alpha}=&i\epsilon_{\alpha}D^a+\left(\sigma^{\mu\nu}\epsilon
\right)_{\alpha}\left[F^a_{\mu\nu}
+\frak12iC_{\mu\nu}d^{abc}\lambdabar^b\lambdabar^c\right]\cr
\delta D^a=&-\epsilon\sigma^{\mu}D_{\mu}\lambdabar^a.\cr}}

Whether the non-anticommutative superfield formalism may be similarly
adapted for $SU(N)$ requires further investigation.\foot{We thank 
S. Terashima for a discussion on this point.}
\newsec{One-loop calculation}
We use the standard gauge-fixing term 
\eqn\gafix{
S_{\rm{gf}}={1\over{2\alpha}}\int d^4x (\pa.A)^2} 
with its associated
ghost terms.  The gauge propagator is  
\eqn\gprop{
\Delta_{\mu\nu}=-{1\over{p^2}}\left(\eta_{\mu\nu}
+(\alpha-1){p_{\mu}p_{\nu}\over{p^2}}\right)}
and the gaugino propagator is  
\eqn\fprop{
\Delta_{\alpha\alphadot}={p_\mu\sigma^{\mu}_{\alpha\alphadot}\over{p^2}},}
where the momentum enters at the end of the propagator with the undotted 
index.  
The one-loop graphs contributing
to the ``standard'' terms in the lagrangian (those without a
$C^{\mu\nu}$) are the same as in the ordinary $N=1$ case, so gauge field
and gaugino anomalous dimensions and gauge $\beta$-function are as for
$N=1$. Since our gauge-fixing term in Eq.~\gafix\ does not preserve 
supersymmetry, the gauge field and gaugino anomalous dimensions are
different (and moreover gauge-parameter dependent). However, the 
gauge $\beta$-function is of course gauge-independent. 
The one-loop one-particle-irreducible (1PI) 
graphs contributing to the new terms (those
containing $C$) are depicted in Figs.~1--3. 
\bigskip
\epsfysize= 2in
\centerline{\epsfbox{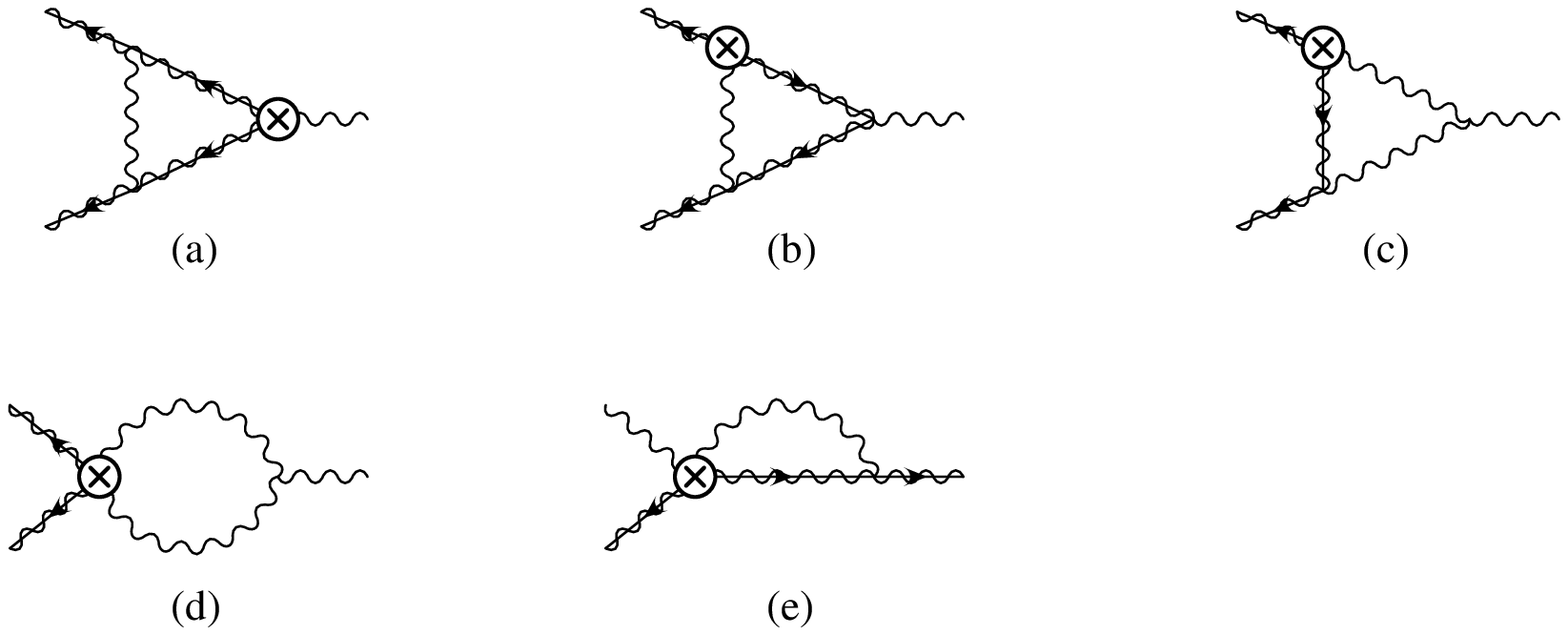}}
\inparg
{\it \noindent Fig.1: Diagrams with one gauge, two gaugino lines; the crossed
circle represents the position of a $C$.}
\medskip
\outparg

The divergent contributions 
to the effective action from the graphs in Fig.~1 are
given by: 
\eqn\grapha{\eqalign{ \Gamma_{1a}^{\rm{pole}}&=-\left(3+\alpha\right)NLT_1\cr
\Gamma_{1b}^{\rm{pole}}&=-NL\left(T_1+\frak43A_1\right)\cr
\Gamma_{1c}^{\rm{pole}}
&=-NL\left[\frak14(2+7\alpha)T_1+\frak13(2+3\alpha)A_1\right]\cr
\Gamma_{1d}^{\rm{pole}}&=\frak12NL(5+\alpha)T_1\cr
\Gamma_{1e}^{\rm{pole}}&=NL\left[\frak14(3-\alpha)T_1+(1+\alpha)A_1\right]\cr}} 
where
\eqn\ttensor{\eqalign{
T_1&=id^{abc}gC^{\mu\nu}\pa_{\mu}A_{\nu}^a\lambdabar^b\lambdabar^c,\cr
A_1&=id^{abc}gC^{\mu}{}_{\nu}A_{\mu}^a\lambdabar^b\sigmabar^{\nu\rho}\pa_{\rho}
\lambdabar^c,\cr}}
and (using dimensional regularisation with $d=4-\epsilon$)
$L={g^2\over{16\pi^2\epsilon}}$. 
\bigskip
\epsfysize= 4in
\centerline{\epsfbox{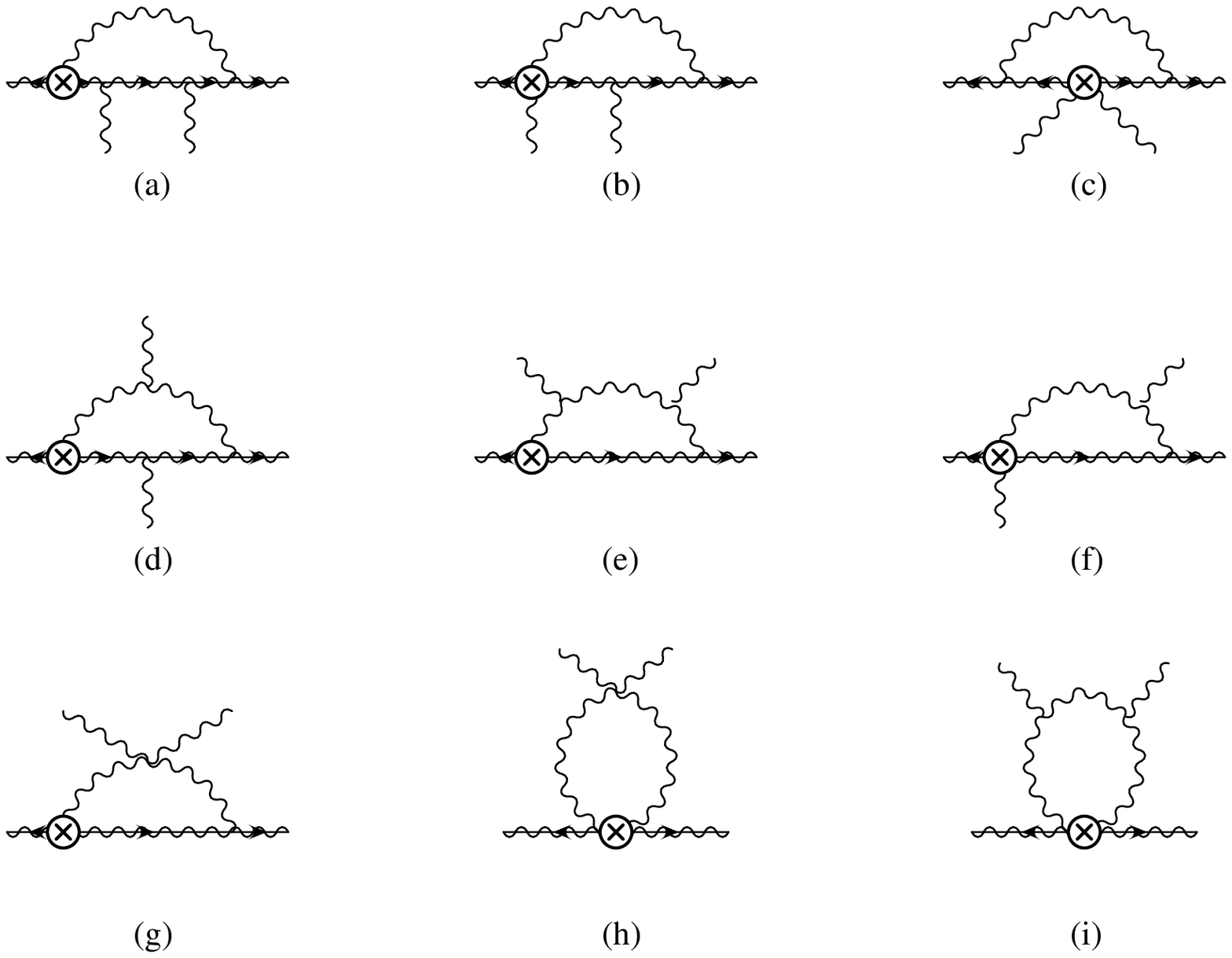}}
\inparg
{\it \noindent Fig.2: Diagrams with two gauge and two gaugino lines; the 
crossed circle represents the position of a $C$.}
\medskip
\outparg

The results for Fig.~2 are given by:
\eqn\graphc{\eqalign{ \Gamma_{2a}^{\rm{pole}}
&=NL\left(\frak12T_2+\frak13A_2\right)\cr
\Gamma_{2b}^{\rm{pole}}&=0\cr 
\Gamma_{2c}^{\rm{pole}}&=\frak12NL(3+\alpha)T_2\cr 
\Gamma_{2d}^{\rm{pole}}&=0\cr
\Gamma_{2e}^{\rm{pole}}
&=NL\left[-\frak12\alpha T_2+\frak16(4+3\alpha)A_2\right]\cr
\Gamma_{2f}^{\rm{pole}}&=NL\left[\frak34\alpha T_2
-\frak12(2+\alpha)A_2\right]\cr
\Gamma_{2g}^{\rm{pole}}&=NL\left(\frak34\alpha T_2+\frak12A_2\right)\cr
\Gamma_{2h}^{\rm{pole}}&=-\frak34NL(1+\alpha)T_2\cr 
\Gamma_{2i}^{\rm{pole}}&=\frak34\alpha NLT_2\cr}}
where 
\eqn\tsecasec{\eqalign{
T_2&=id^{abe}f^{cde}g^2C^{\mu\nu}A_{\mu}^cA_{\nu}^d\lambdabar^a\lambdabar^b,\cr
A_2&=id^{cde}f^{abe}g^2C^{\mu\rho}A_{\mu}^cA_{\nu}^d
\lambdabar^a\sigmabar^{\nu\rho}\lambdabar^b.\cr}} 
\bigskip
\epsfysize= 2in
\centerline{\epsfbox{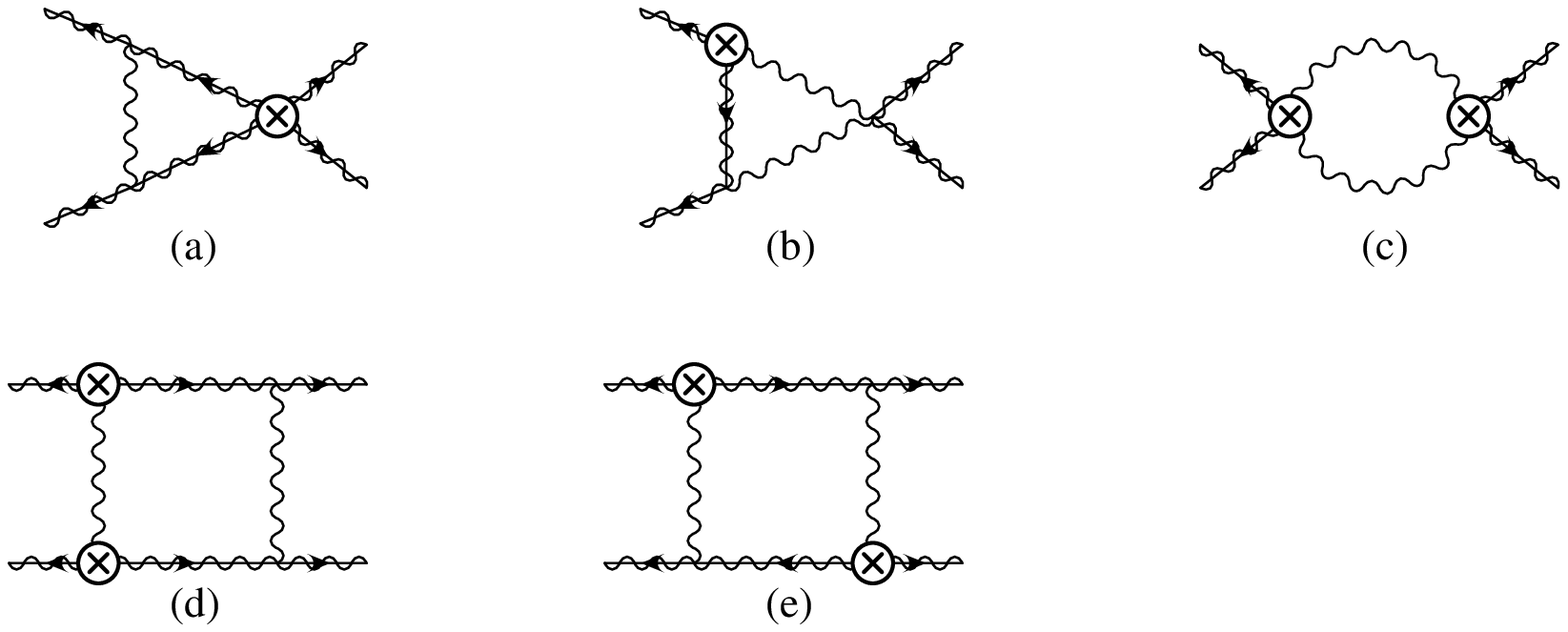}}
\inparg
{\it \noindent Fig.3: Diagrams with four gaugino lines; the crossed 
circle represents the position of a $C$ or $|C|^2$.}
\medskip
\outparg

Finally, the results
for Fig.~3 are given by: 
\eqn\graphb{\eqalign{
\Gamma_{3a}^{\rm{pole}}&=L\Bigl[\frak14(3+\alpha)NT_3+h(3+\alpha)\Ttil_3
-\frak4Nf^{abe}f^{cde}hg^2|C|^2(\lambdabar^a\lambdabar^c)
(\lambdabar^b\lambdabar^d)\cr
&+2d^{abcd}g^2|C|^2[(\lambdabar^a\lambdabar^b)
(\lambdabar^c\lambdabar^d)
-\frak12(\alpha-1)(\lambdabar^a\lambdabar^d)(\lambdabar^c\lambdabar^b)]
\Bigr]\cr
\Gamma_{3b}^{\rm{pole}}&=\frak12\alpha NLT_3\cr 
\Gamma_{3c}^{\rm{pole}}&=-\frak14(1+\alpha)NLT_3\cr
\Gamma_{3d}^{\rm{pole}}&=-2d^{abcd}Lg^2|C|^2 
[(\lambdabar^a\lambdabar^b)(\lambdabar^c\lambdabar^d)
-\frak12(\alpha-1)(\lambdabar^a\lambdabar^d)(\lambdabar^c\lambdabar^b)]\cr
\Gamma_{3e}^{\rm{pole}}&=\frak13f^{efa}f^{ghb}d^{gfc}d^{ehd}Lg^2|C|^2[
(\lambdabar^a\lambdabar^b)(\lambdabar^c\lambdabar^d)   
-(\lambdabar^a\lambdabar^d)(\lambdabar^c\lambdabar^b)]\cr}} 
where
\eqn\dtensor{\eqalign{ 
T_3=&d^{abe}d^{cde}g^2|C|^2
(\lambdabar^a\lambdabar^b)(\lambdabar^c\lambdabar^d),\cr
\Ttil_3=&g^2|C|^2(\lambdabar^a\lambdabar^a)(\lambdabar^b\lambdabar^b),\cr
d^{abcd}=&f^{iaj}f^{jbk}d^{kcl}d^{ldi}.\cr}}
In obtaining these results we have made frequent use of the identity
\eqn\fdida{
d^{iaj}f^{jbk}f^{kci}=-{N\over 2} d^{abc}.}
Other useful identities are
\eqn\fdidb{\eqalign{
f^{abe}d^{ecd}+f^{ace}d^{edb}+f^{ade}d^{ebc}&=0,\cr
f^{abe}f^{cde}&=\frak2N(\delta^{ac}\delta^{bd}-\delta^{ad}\delta^{bc})
+d^{ace}d^{bde}-d^{ade}d^{bce}.\cr}}
Using\ref\azcar{J.A.~de~Azcarraga,
A.J.~Macfarlane, A.J.~Mountain and J.C.~Perez Bueno, \npb510 (1998) 657}
 \eqn\fdid{ f^{efa}f^{ghb}d^{gfc}d^{ehd}={N\over
4}(d^{abf}d^{cdf}-d^{adf}d^{bcf}- d^{acf}d^{bdf}),}
together with the Fierz identity
\eqn\fierz{
(\lambdabar^a\lambdabar^b)(\lambdabar^c\lambdabar^d)
+(\lambdabar^a\lambdabar^c)(\lambdabar^b\lambdabar^d)
+(\lambdabar^a\lambdabar^d)(\lambdabar^b\lambdabar^c)=0}
we find
\eqn\newG{
\Gamma_{3e}^{\rm{pole}}=\frak14NLT_3.}
Then adding Eqs.~\grapha, \graphc\ and \graphb\ we find the full
result for the divergent contribution to the 
one-loop effective action from diagrams with one or two $C$s: 
\eqn\fullres{\eqalign{
\Gamma_{\rm{1PI}}^{(1)\rm{pole}}
=&NL\int d^4x\Bigl[
-\frak54(1+2\alpha)T_1+\frak14(5+6\alpha)T_2+\frak12\left(\frak32+\alpha
-\frak{12}{N^2}h\right)T_3
\cr
&+\frak{h}{N}\left(3+\alpha-\frak{12}{N^2}\right)\Ttil_3
-A_1+\frak12A_2\Bigr].\cr}} 
These results differ from those of Ref.~\alish\ (even making the
natural allowances
for their use of the gauge group $U(N)$). The most
obvious difference is that they claim that several diagrams, for which
we obtain non-zero results, vanish when the diagrams with ``crossed''
fermion lines are taken into account.  For instance, they claim that
Figs.~1(c)--(e) vanish identically. This  conclusion seems manifestly
false; it is easiest to see this in the case of  Fig.~1(d) where there
are only external gaugino lines. Moreover the identity
$C^{\rho\nu}\sigmabar_{\nu}^{\betadot\delta}\sigma^{\kappa}
_{\delta\deltadot}=-C^{\rho\kappa}\delta^{\betadot}_{\deltadot}$ (below
their  Eq.~(45)) is simply not true.  
 
At first sight, though, our results appear implausible; the terms $A_1$ and 
$A_2$ are clearly 
problematic since they violate gauge invariance. However,  
we have found that remarkably this difficulty can be resolved and the 
theory rendered renormalisable at a stroke by a divergent redefinition
\eqn\lchange{
\delta\lambda^a=-\frak12NLgC^{\mu\nu}d^{abc}
\sigma_{\mu}\lambdabar^cA_{\nu}^b.}
Note that this only affects the gaugino kinetic term. 
This results in a change in the action
\eqn\delS{
\delta S=NL\int d^4x\left[-\frak14(T_1-T_2)+A_1
-\frak12A_2\right].}
We therefore have 
\eqn\delsum{\eqalign{
\Gamma_{\rm{1PI}}^{(1)\rm{pole}\prime}
=\Gamma_{\rm{1PI}}^{(1)\rm{pole}}+\delta S
=&NL\int d^4x
\Bigl[-\frak12(3+5\alpha)T_1+
\frak32(1+\alpha)T_2\cr
&+\frak12\left(\frak32+\alpha-\frak{12}{N^2}h\right)T_3
+\frak{h}{N}\left(3+\alpha-\frak{12}{N^2}\right)\Ttil_3.\Bigr].\cr}}
If we introduce 
bare fields and couplings according to    
\eqn\bare{\eqalign{ \lambdabar_B=Z_{\lambda}^{\frak12}\lambdabar,\quad
A_{B\mu}=&Z_A^{\frak12}A_{\mu},\quad g_B=Z_gg, \cr
C_B=Z_CC, \quad|C|_B^2=&Z_{|C|^2}|C|^2,\quad h_B=Z_h h\cr}} 
the $C$-dependent part of the bare action can be written as
\eqn\dgtnew{\eqalign{ S_B=&\int
d^4x\Bigl[-Z_CZ_gZ_A^{\frak12}Z_{\lambda}T_1+\frak12Z_CZ_g^2Z_AZ_{\lambda}T_2
\cr
&+\frak18Z_{|C|^2}Z_g^2Z_{\lambda}^2T_3
+\frak14Z_{|C|^2}Z_g^2Z_{\lambda}^2Z_h\frak{h}{N}\Ttil_3\Bigr].\cr}}
Of course a natural expectation would be that $Z_{|C|^2}=(Z_C)^2$ but
this cannot be assumed as yet. As we mentioned before,
the renormalisation constants for the fields
and for the gauge coupling $g$ are the same as in the ordinary $N=1$
supersymmetric theory\lunin\ and are therefore given by\ref\timj{
D.~Gross and F.~Wilcek, \prd8 (1973) 3633\semi
D.R.T.~Jones, \npb87 (1975) 127}: 
\eqn\Zgg{\eqalign{
Z_{\lambda}&=1-2\alpha NL,\cr
Z_A&=1+(3-\alpha)NL\cr
Z_g&=1-3NL.\cr}}
Note that $h$ is multiplicatively renormalised since the 
divergent term in $\Ttil_3$ is proportional to $h$, the coefficient
of $\Ttil_3$ in the lagrangian Eq.~\lagran; this is due to the 
cancellation of the $d^{abcd}$ terms in Eq.~\graphb\ between 
$\Gamma_{3a}^{\rm{pole}}$ and $\Gamma_{3d}^{\rm{pole}}$ 
which otherwise would have produced non-homogeneous terms. This will
be important shortly.
We now find that if we take
\eqn\Zres{\eqalign{
Z_C=1, \quad &Z_{|C|^2}=1+48\frak{h}{N}L,\cr
Z_h=&1-6NL\left[1+8{(h-1)\over{N^2}}\right],\cr}}
then
\eqn\gammone{
\Gamma^{(1)\rm{pole}\prime}=\Gamma_{\rm{1PI}}^{(1)\rm{pole}\prime}+S_B^{(1)}=0,}
i.e. $\Gamma^{(1)\prime}$ is finite. 
So the form of the lagrangian is not quite preserved under renormalisation 
due to the non-zero $h$ and the fact that
$Z_h\ne1$. As we remarked earlier, the original
$N=\frak12$ supersymmetric action\seib\ corresponds to taking $h=1$ in 
Eq.~\lagran.
However because $h$ is multiplicatively renormalised and $Z_{|C|^2}-1\propto h$,
 if we set $h=0$ in
Eq.~\lagran\ then we obtain a lagrangian whose form {\it is} 
exactly preserved under renormalisation--without even any need to renormalise
the anticommutativity parameter $C$. The lagrangian with $h=0$ is still
$N=\frak12$ supersymmetric and of course gauge-invariant.
\newsec{Conclusions}
We have shown how a combination of a minor modification to the 
pure $N=\frak12$ supersymmetric action and a 
gauge-non-invariant divergent field redefinition lead to an action whose form 
is preserved under renormalisation at one loop. This is surprising because
with this relatively simple action, $N=\frak12$ supersymmetry imposes no  
constraints so we could have expected $Z_C$ and $Z_{|C|^2}$ to be different and
not equal to unity; (though with $Z_C\ne1$, $C$ would have to be replaced by 
$C_B$ in the $N=\frak12$ supersymmetry transformations). It will be 
interesting to see what happens in the case of $N=\frak12$ supersymmetry 
including chiral matter, where there are several terms involving $C$ and
linked by $N=\frak12$ supersymmetry. Work on
this is in progress and we expect to report on it shortly. Of course 
it will also be worthwhile to see whether these properties persist
at higher loops.

We have restored gauge invariance in this case by a somewhat unconventional 
expedient which works rather miraculously. One could speculate to 
what extent the $N=\frak12$ supersymmetry and the identities Eq.~\cprop\
were required to make this trick work.
It would be interesting to examine a theory of the same form but in which 
$C^{\mu\nu}$ was replaced by a general antisymmetric tensor. Moreover,
suppose one considered a theory with an action based on Eq.~\lagran\ but
including all the hermitian conjugate terms which are missing. The only new 
diagrams would simply be the ``hermitian  conjugates'' of those in 
Figs.~1--3. Eq.~\lchange\ would now need to be supplemented by its hermitian
conjugate. However, the variation of the action would now include
additional unwanted non-gauge-invariant terms since it is now not only the 
gaugino kinetic term which varies. This raises the spectre of a theory 
(albeit non-renormalisable) with ineradicable non-gauge-invariant 
divergences.

We have adapted the $N=\frak12$ action in components from the gauge group 
$U(N)$ to $SU(N)$, partly because this simplifies the 
calculations. On the other hand,
from our current perspective $U(N)$ presents the {\it prima facie}
advantage that the $\Ttil_3$ term would not appear explicitly in Eq.~\lagran\ 
and
one might hope that Eq.~\lagran\ might be form-invariant under renormalisation
as it stands. However, the $U(N)$ version of Eq.~\fdid\ includes extra
terms\ref\bonora{L.~Bonora and M.~Salizzoni, \plb504 (2001) 80}
which lead to $\Ttil_3$ being generated despite not being present in
Eq.~\lagran--so the form-invariance is inescapably lost in this case.
Moreover, while the $SU(N)$ gauge coupling renormalises in accordance with
Eq.~\Zgg, the $U(1)$ coupling would of course be unrenormalised.  
  
Finally we do not as yet have a theoretical justification or interpretation for 
the field redefinition which we appear to be compelled to make, and this point
deserves investigation.
  
\bigskip\centerline{{\bf Acknowledgements}}\nobreak

DRTJ was supported by a PPARC Senior Fellowship, and a CERN Research 
Associateship, and was visiting CERN while most of this work was done.
LAW was supported by PPARC through a Graduate Studentship. One of us
(DRTJ) thanks Luis Alvarez-Gaum\'e for a conversation. 

\listrefs
\bye